\documentclass[prl,superscriptaddress,twocolumn, noshowpacs, amsmath,amssymb]{revtex4}

\usepackage{graphicx}
\usepackage{dcolumn}
\usepackage{bm}
\usepackage{color}
\usepackage{mathrsfs}

\usepackage{xcolor}

\begin{document}


\title{Helical spin texture and interference of Majorana bound states in one-dimensional topological superconductor}

\author{Takuto Kawakami}
\email{KAWAKAMI.Takuto@nims.go.jp}
\affiliation{International Center for Materials Nanoarchitectonics (WPI-MANA),
National Institute for Materials Science, Tsukuba 305-0044, Japan}

\author{Xiao Hu}
\email[Correspondence to: ]{HU.Xiao@nims.go.jp}
\affiliation{International Center for Materials Nanoarchitectonics (WPI-MANA),
National Institute for Materials Science, Tsukuba 305-0044, Japan}


\date{\today}

\begin{abstract}{We investigate one-dimensional (1D) Majorana bound states (MBSs) realized in terms of the helical edge states of a 2D quantum spin-Hall insulator (QSHI) in a
heterostructure with a superconducting substrate and two ferromagnetic insulators (FIs).
By means of Bogoliubov-de Gennes approach we demonstrate that there is a helical spin texture in the MBS wave function with a pitch proportional to the
Fermi momentum of the helical edge states of QSHI
. Moreover, simultaneous detection on local density of states by scanning tunneling microscopy and spectroscopy at a position close to one FI edge and at the midpoint
between two FIs can not only map out the energy spectrum $\pm E \cos(\phi/2)$, but also prove experimentally that the two quasiparticle excitations do not mix
with each other as protected by the parity conservation associated with the MBSs.}
\end{abstract}

\maketitle

{\it Introduction.---}
Considerable efforts have been devoted to search for peculiar zero-energy quasiparticle excitations in topological superconductors
(SCs)~\cite{read_green,kitaev,hasan2010,qi2011}, which, getting name of Majorana bound states (MBSs)
due to the equivalence to their antiparticles~\cite{majorana,wilczek}, can realize non-Abelian quantum statistics useful for
establishing decoherence-free topological quantum computation~\cite{ivanov2001,nayak2008,alicea2011,alicearpp,beenakkerarcp,liang_hu,wu_hu}.
While it becomes clear theoretically that
MBSs appear at vortex cores in two-dimensional (2D) topological SCs and ends of 1D ones realized in various hybrid
systems~\cite{fu2008,fu2009,sato2009,lutchyn2010,sau2010,jiang2013,hui2015,mourik2012,deng2012,finck2013},
to nail them down conclusively in experiments remains challenging~\cite{alicearpp,wang_hu,he2014,liu2015}.

Recently experimental works have been reported on local density of states (LDOS) associated with MBSs explored by scanning tunneling microscopy and spectroscopy (STM/STS).
In a system of iron (Fe) chains on the lead (Pb) substrate, high-resolution spectroscopic images signal zero-energy quasiparticle states localized at the ends
of Fe chains, as expected with the emergence of MBSs~\cite{Yazdani}.
In a hybrid device made of a NbSe$_2$ substrate and a thin film of 3D topological insulator (TI) Bi$_2$Te$_3$, LDOS at vortex cores
as a function of energy and the distance from vortex center evolves from ``V''-shape to ``Y''-shape with increasing thickness of the TI film, reflecting the appearance of MBSs~\cite{xu2015}.
By means of Bogoliubov-de Gennes (BdG) approach the present authors confirmed this LDOS evolution, and meanwhile
revealed a ``checkerboard'' pattern in the relative LDOS between spin up and down channels inside vortex cores. The latter property
can be explored experimentally by spin-resolved STM/STS, which may identify MBSs as individual quantum states~\cite{kawakami2015}.

\begin{figure}[t]
\includegraphics[width=8cm]{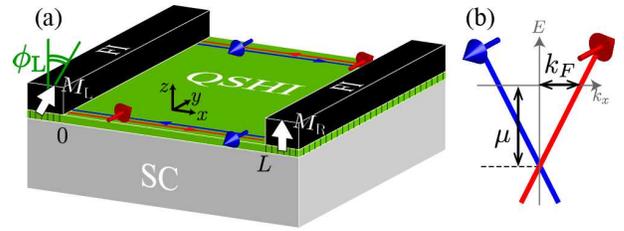}
\caption{ System geometry with a 2D quantum spin-Hall insulator on top of a superconductor substrate with two ferromagnetic insulators attached.
}
\label{fig:model}
\end{figure}

The spin oscillations in MBS wave function in a vortex core are governed by the conservation of total angular momentum, which has contributions from orbital and spin
angular momenta as well as the phase winding of superconducting vortex. It is then interesting to ask whether a spin texture exists in 1D MBS where orbital
angular momentum and superconducting phase winding cannot be defined. In this regard, we notice that very recently novel 1D superconducting states were reported
in hybrid systems of HgTe/HgCdTe~\cite{Molenkamp} and InAs/GaSb~\cite{pribiag2015} quantum wells in proximity to SCs, a useful platform proposed theoretically
some time ago for realizing effective $p$-wave topological superconducting state in terms of the helical edge states of quantum spin Hall insulator (QSHI)
and $s$-wave SC~\cite{fu2009}. Attaching ferromagnetic insulator (FI) additionally MBS can be realized at the edge of FI~\cite{fu2009}, with wave functions derived analytically
and fractional Josephson-like current-phase relation addressed under general conditions~\cite{jiang2013}.

In this work, with the recent progresses in STM/STS techniques in mind we revisit this problem paying special attentions to spin textures and LDOS of MBSs.
Analyzing the system schematically shown in Fig.~\ref{fig:model} in terms of BdG approach,
we demonstrate a 1D helical spin texture in MBS with a pitch proportional to the Fermi momentum
of the topological edge states of QSHI, which can be tuned by rotating the magnetization direction $\phi$ of one FI relatively to the other.
Moreover, simultaneous detection on LDOS at a position close to one FI edge and at the midpoint between two FIs can not only map out
the energy spectrum $\pm E \cos(\phi/2)$, but also prove experimentally that the two quasiparticle excitations do not mix with each other as protected by the
parity conservation associated with the MBSs.

\vspace{3mm}
{\it BdG Hamiltonian and spin texture of MBS}.--- The low-energy physics of the
system depicted schematically in Fig.~\ref{fig:model} is described by the following
1D BdG equation
\begin{eqnarray}\label{eq:H1D} 
\left[\begin{array}{cc}
h_0\!+\!\bm{M}\!\cdot\!\hat{\bm s} & -i\Delta_0 \hat{s}_y \\
i\Delta_0 \hat{s}_y & -\!h_0^\ast\!-\!\bm{M}\!\cdot\!\hat{\bm s}^{\ast} \end{array}\right]
\left(\!\begin{array}{c}
\vec{u}_E \\
\vec{v}_E
\end{array}\!\right)
\!=\!E\left(\!\begin{array}{c}
\vec{u}_E \\
\vec{v}_E
\end{array}\!\right)
\end{eqnarray}
with $h_0$ for the helical edge states of QSHI
\begin{eqnarray}\label{eq:hdirac}
h_0(x) = -\mu -iv_{F}\partial_x \hat{s}_y,
\end{eqnarray}
where the spin operators $\hat{\bm s}$ standing for $2\times2$ Pauli matrices, and the Nambu spinor is given by $\vec{u}_E=(u_{E}^{\uparrow},u_{E}^{\downarrow})^T$
and $\vec{v}_E=(v_{E}^{\uparrow},v_{E}^{\downarrow})^T$ with $\uparrow\downarrow$ denoting spin up and down states.
To be specific, chemical potential $\mu$ is measured from the Dirac point of topological edge states of QSHI, and $\mu>0$ and $v_{\rm F}>0$ is considered.
The spin polarization in the helical edge states of QSHI is taken in the way such that the right and left moving
quasiparticles carry $+y$ and $-y$ spin respectively (see Fig.~\ref{fig:model}).
Magnetizations finite outside the region $[0, L]$ are confined in the $xz$ plane, which for simplicity are
assigned the same strength $M\equiv |\bm{M}_{\mathrm L}|=|\bm{M}_{\mathrm R}|$.

For sufficiently large $L$, zero-energy MBSs can be generated at the edges of FIs when the Zeeman energy satisfies the condition $M>\sqrt{\mu^2+\Delta_0^2}$~\cite{fu2009}.
In a form with spinors and rotation operators given explicitly (in contrary to the previous work \cite{jiang2013}), one has the MBS wave functions
\begin{eqnarray}
\vec{u}_0^{L}\!=\!\vec{v}_0^{L\ast}\!=\!
\left\{\!
\begin{array}{l}
e^{-\frac{x}{\xi_{\mathrm{S}}}} e^{i\hat{s}_y\left(xk_{\mathrm{F}}-\frac{\theta_{\mathrm L}}{2}\right)}
\!\left(\!\begin{array}{c}
1 \\
0 \\
\end{array}\!\right), \ \hbox{for}\ x\ge 0; \\
e^{\frac{x}{\xi_{\mathrm{M}}}}
e^{-i\hat{s}_y \theta_{\mathrm L}/2}
\left(\!\begin{array}{c}
1 \\
0
\end{array}\!\right), \ \hbox{for}\ x<0,
\end{array}\right.\label{eq:MBSL}
\end{eqnarray}
and
\begin{eqnarray}\label{eq:MBSR}
\vec{u}_0^{R}\!=\!\vec{v}_0^{R\ast}\!=\!
\left\{\!
\begin{array}{l}
ie^{\frac{x\!-\!L}{\xi_{\mathrm{S}}}}\!e^{i\hat{s}_y\!\left[\!(\!x\!-\!L\!)k_{\mathrm{F}}\!-\!\frac{\theta_{\mathrm R}}{2}\!\right]}
\!\left(\!\begin{array}{c}
1 \\
0 \\
\end{array}\!\right)\!,\hbox{ for } x\le L; \!\\
ie^{-\frac{x\!-\!L}{\xi_{\mathrm{M}}}}
e^{-i\hat{s}_y\frac{\theta_{\mathrm R}}{2}}
\left(\!\begin{array}{c}
1 \\
0
\end{array}\!\right), \ \hbox{for}\ x>L,\!
\end{array}\!\right.\!
\end{eqnarray}
with $\theta_{\mathrm L/R} = \phi_{\mathrm L/R} \pm 2\arctan[\sqrt{(M-\mu)/(M+\mu)}]$,
$\phi_{\mathrm L/R}$ the magnetization direction of the left/right FI (see Fig.~\ref{fig:model}),
$\xi_{\mathrm{S}}=v_{\mathrm F}/\Delta_0$ and  $\xi_{\mathrm{M}}=v_{\mathrm F}/(\sqrt{M^2-\mu^2}-\Delta_0)$.

\begin{figure}[t]
 \includegraphics[width=80mm]{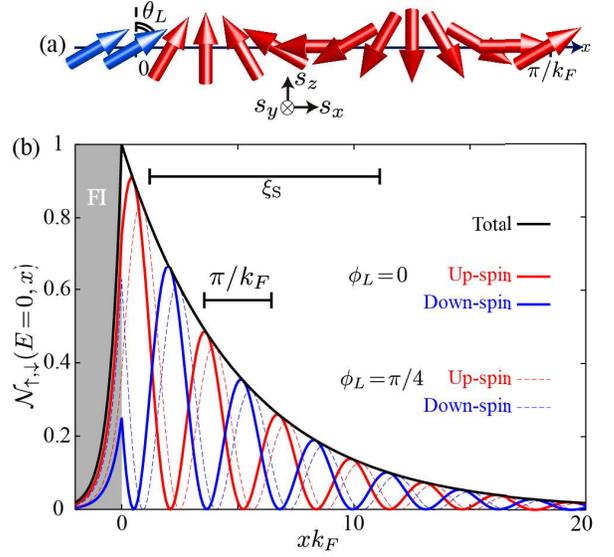}
 \caption{(a) Helical spin texture in MBS with blue and red arrows denoting the spin direction in FI region and TI region
 respectively, and (b) spin-resolved LDOS associated with MBS for $\phi_{\mathrm L}=0$ and $\phi_{\mathrm L}=\pi/4$,
 with the length given in units of $k_{\mathrm F}^{-1}=v_F/\mu$. LDOS are normalized by the total DOS at the edge of FI.
 Parameters are taken as $\mu=10 \Delta_0$ and $M=20\Delta_0$ for which $\xi_{\mathrm{S}} k_{\rm F}=10$.
}
\label{fig:helicalspin}
\end{figure}

The spinor $(1,0)^T$ in Eqs.~(\ref{eq:MBSL}) and (\ref{eq:MBSR}) is the eigenstate of spin operator $\hat{s}_z$ and $\exp(i\hat{s}_y\gamma)$ is the
rotation operator around the $y$-axis by angle $2\gamma$. Since in Eq.~(\ref{eq:MBSL}) the angle
is linearly proportional to the coordinate for $x\ge 0$ [whereas for $x\le L$ in Eq.~(\ref{eq:MBSR})], the MBS wave function exhibits a helical spin structure
in the $xz$ plane with the pitch $k_{\mathrm F}/\pi$, which is shown explicitly in Fig.~\ref{fig:helicalspin}(a) in terms of the local spin moment
$\bm{S}(x) = \frac{1}{|\vec{u}_0(x)|^2} \vec{u}_{0}^\dag(x) \hat{\bm s} \vec{u}_{0}(x)$. The helical spin texture of MBS can be
shifted along the $x$ direction by rotating the magnetization direction of FI $\phi_{\rm L}$ or $\phi_{\rm R}$. The wave functions of MBSs in Eqs .~(\ref{eq:MBSL}) and (\ref{eq:MBSR})
are real and purely imaginary respectively because the spin polarization in the helical edge states of QSHI is taken along the $y$ axis and the magnetization
of FI is confined in the $xz$ plane.

Experimentally spin textures in quasiparticle excitations can be measured by spin-resolved STM/STS~\cite{bode2003, Yazdani}, where the norms of wave functions
are measured by $dI/dV$ in spin-up and spin-down channels separately
\begin{eqnarray}
\mathcal{N}_{s}(E,x)=\sum_{E'}|u^{s}_{E'}(x)|^2\delta(E'-E)
\label{eq:ldos}
\end{eqnarray}
with $s=\uparrow\downarrow$.
As shown in Fig.~\ref{fig:helicalspin}(b), LDOS in spin-up and -down channels derived from Eq.~(\ref{eq:MBSL}) exhibit a quantum oscillations.
In HgTe/HgCdTe quantum wells, one has $v_F\simeq 2.4$ \AA$\cdot$eV~\cite{qi2011}.
For a typical chemical potential $\mu=4$ meV, the
wave length of helical spin texture is estimated as $\pi/k_F=188$ nm, which can be well resolved by the state-of-art spin-resolved STM/STS.

\begin{figure}[t]
 \includegraphics[width=80mm]{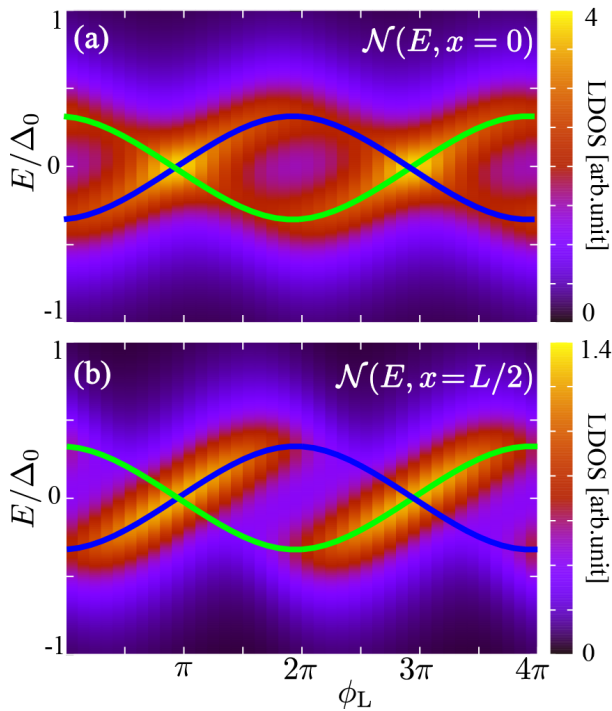}
 \caption{Energy spectrum and LDOS at the end (a) and midpoint (b) of the system as a function of the magnetization direction of the left FI $\phi_{\rm L}$.
  Parameters are the same as Fig.~\ref{fig:helicalspin} except for $L=1.83\xi_{\mathrm{S}}$, $\eta=0.25\Delta_0$ and $\phi_{\mathrm R}=0$.
  }
\label{fig:LDOScos}
\end{figure}

\vspace{3mm}
{\it Interference between two MBSs.---}
When the system size is comparable with the length $\xi_{\rm S}$, MBSs at the two ends interact with each other due to the interference between their wave functions.
In terms of perturbation approximation on Hamiltonian~(\ref{eq:H1D}) with the two MBS wave functions given in Eqs.~(\ref{eq:MBSL}) and (\ref{eq:MBSR}), one derives the energy spectrum
as (see also \cite{jiang2013})
\begin{equation}
E=\pm E_0\sin[(\theta_{\mathrm L} - \theta_{\mathrm R})/2-Lk_F],
\label{eq:spectrum}
\end{equation}
with
\begin{equation}
E_0=2\Delta_0 e^{-L/\xi_{\mathrm{S}}}\left(1-\frac{\Delta_0}{\sqrt{M^2-\mu^2}}\right).
\label{eq:amplitude}
\end{equation}
The associated wave functions of the two quasiparticle excitations are given by
\begin{eqnarray}
\left(\begin{array}{c}
\vec{u}_{\pm} \\
\vec{v}_{\pm}
\end{array}\right)
=
\left(\begin{array}{c}
\vec{u}_0^{\mathrm L}\\
\vec{v}_0^{\mathrm L}
\end{array}\right)
\mp i
\left(\begin{array}{c}
\vec{u}^{\mathrm{R}}_0\\
\vec{v}^{\mathrm{R}}_0
\end{array}\right),
\label{eq:bonding}
\end{eqnarray}
namely the bonding and antibonding states formed by the two end MBSs, corresponding to the even and
odd parity states of the composed fermion (note that generally $\vec{u}_{\pm}\neq \vec{v}^{\ast}_{\pm}$).

Theoretically it is clear that there is no gap opening when the two energy levels cross each other at zero energy
since the parity is conserved in the present system. However, to prove this experimentally requires precise
detection on level crossing, which is generally difficult even using the
state-of-art STM technique since reduction of energy resolution is inevitable due to operation at finite temperatures.
This can be seen in Fig.~\ref{fig:LDOScos}(a) where LDOS close to the left FI obtained by  numerically solving BdG equation (\ref{eq:H1D})
is shown with $\delta(E)$ function in Eq.~(\ref{eq:ldos}) replaced by a smearing function $C(E,\eta)=E/[\pi(\eta^2+E^2)]$ with $\eta=0.25\Delta_0$ (corresponding to operation temperature
of 1 K when a SC gap $\Delta_0=0.4$ meV is presumed).

\begin{figure}[t]
\includegraphics[width=80mm]{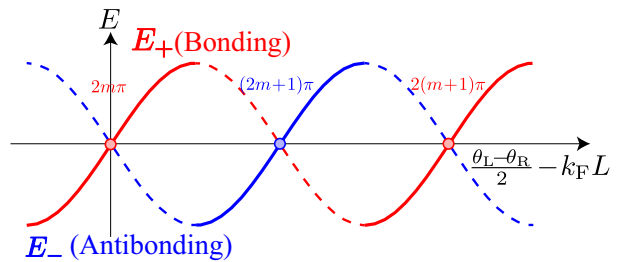}
\caption{Schematics for energy spectrum Eq.~(\ref{eq:spectrum}) with LDOS at the midpoint between the two MBSs given in Eq.~(\ref{eq:qpmidpoint}),
with sold and dashed parts for wave functions enhanced and suppressed respectively}.
\label{fig:EDOS}
\end{figure}

We notice, however, that in the present system one can overcome this difficulty by tracing LDOS at $x=L/2$,
the midpoint of the QSHI edge where the two MBS wave functions take the same amplitude, as a function of $\phi_{\rm L}$.
In order to demonstrate this point, we present explicitly the wave functions of quasiparticle excitations at $x=L/2$:
\begin{eqnarray}\label{eq:qpmidpoint}
\vec{u}_{\pm}(x=L/2)=
\left[ 1 \pm e^{i\hat{s}_y(\frac{\theta_L-\theta_R}{2}-k_\mathrm{F}L)}\right]\vec{u}_{0}^{\mathrm L}(x=L/2).
\end{eqnarray}
It is then clear that around $(\theta_{\rm L}-\theta_{\rm R})/2-k_{\mathrm F}L = 2m\pi$ with $m$ an integer, the bonding state takes a finite amplitude 
while that of the antibonding state is suppressed to zero as shown schematically in Fig.~\ref{fig:EDOS}. Vice versa,
around $(\theta_{\rm L}-\theta_{\rm R})/2-k_{\mathrm F}L = (2m+1)\pi$ the antibonding state exhibits a finite amplitude while that of the bonding state is suppressed to zero.
This explains the results shown in Fig.~\ref{fig:LDOScos}(b).
Since LDOS in one of the two quasiparticle excitations is zero when the two energy levels cross each other,
mixing between them is obviously impossible. Experimentally, simultaneous detection on LDOS by STM/STS at a position close to FI and at the midpoint
of between the two FIs as shown in Fig.~\ref{fig:LDOScos} will provide a clear evidence for the $4\pi$-periodic energy spectrum associated with the MBSs.

\begin{figure}[b]
\begin{center}
 \includegraphics[width=70mm]{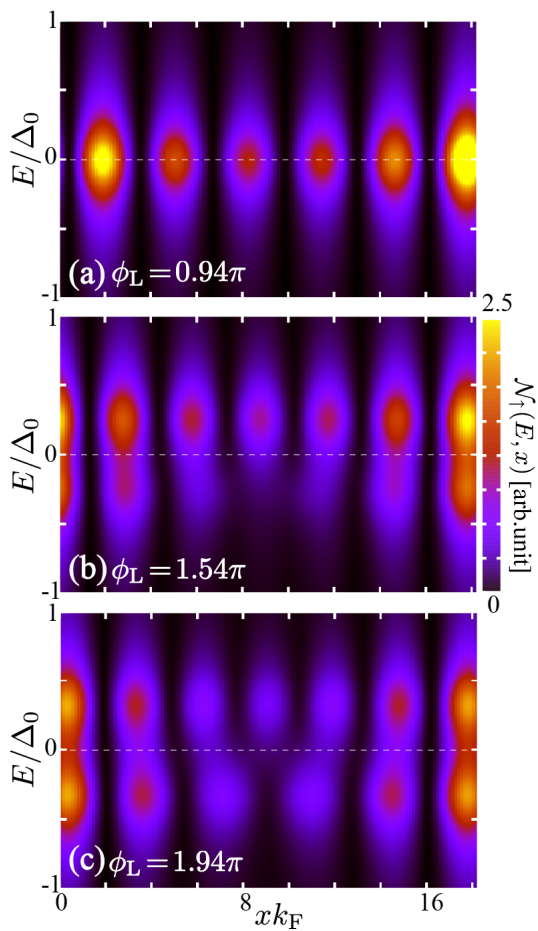}
 \caption{LDOS in spin-up channel for typical values of $\phi_{\rm L}$.
  Parameters are the same as Fig.~\ref{fig:LDOScos}.
  }
\label{fig:LDOSup}
\end{center}
\end{figure}

At the midpoint of between the two FIs, the quasiparticle with finite LDOS always carries energy increasing with $\phi_L$ as seen in Fig.~\ref{fig:LDOScos}(b)
(see also Fig.~\ref{fig:EDOS}). This result is for a positive Fermi velocity ($v_{\rm F}>0$) in the helical edge states of QSHI as given by
Eq.~(\ref{eq:hdirac}). For $v_{\rm F}<0$, the quasiparticle with finite LDOS should carry energy decreasing with $\phi_L$,
a behavior opposite to that in Fig.~\ref{fig:LDOScos}(b), since the energy spectrum of quasiparticle excitation is governed by the helical
edge states of QSHI and magnetization in the way $(\bm{v}\times \bm{s})\cdot \bm{M}$, where the
cross product reverses its sign with $v_{\rm F}$.
Therefore, one can determine the sign of Fermi velocity, an intrinsic
property of the topological edge states of QSHI~\cite{Murakami}, by measuring LDOS at the midpoint
of between the two FIs as shown in Fig.~\ref{fig:model}.

From Eq.~(\ref{eq:spectrum}) it is clear that energies of quasiparticle excitations in a finite
system are dominated by interference between the spin textures generated from the two FIs. This
feature can be revealed more directly by mapping out LDOS in one of the two spin channels, which can be detected experimentally
by spin-resolved STM/STS. As displayed in
Fig.~\ref{fig:LDOSup}(a), MBSs survive when $\phi_{\rm L}$ generates to an integer number of
waves of helical spin texture inside the system. When $\phi_{\rm L}$ increases from this matching
value, another wave enters gradually into the system from the left end which pushes
all the other waves to high energies [see Fig.~\ref{fig:LDOSup}(b)], and finally kicks out that at the midpoint of system [see Fig.~\ref{fig:LDOSup}(c)]. 
A similar behavior can be seen for LDOS in spin-down channel since the
total LDOS exhibits a smooth curve as given in Fig.~\ref{fig:helicalspin}.

\vspace{3mm}
{\it Conclusions.---}
We revisit the Majorana bound states in a heterostructure of two-dimensional quantum spin-Hall insulator and
a $s$-wave superconductor with two ferromagnetic insulators attached.
In terms of Bogoliubov-de Gennes formalism, we demonstrate a helical spin texture in the Majorana-bound-state wave function
with a pitch proportional to the Fermi momentum of helical edge states in quantum spin Hall insulator.
In finite samples the two quasiparticle excitations generated due to interference between Majorana bound states do not mix with each other
due to parity conservation, as can be seen in terms of the local density of states manipulated by
the relative magnetization direction between two ferromagnetic insulators.  
These properties can be explored experimentally by scanning tunneling microscopy and spectroscopy, which provides strong evidences for the 
existence of Majorana bound states.

This work was supported by the WPI Initiative on Materials Nanoarchitectonics, Ministry of Education, Culture, Sports, Science and Technology of Japan.




\end{document}